\documentclass[12pt]{article}
\usepackage{amssymb}
\usepackage{epsfig}

\newcommand{\beqn}{\begin{eqnarray}}
\newcommand{\eeqn}{\end{eqnarray}}

\newcommand{\dd}{{\mbox d}}
\newcommand{\cL}{{\cal L}}
\newcommand{\eq}[1]{(\ref{#1})}

\newcommand{\itep}
{~\vspace{-1.5cm}
\begin{flushright}
{\large ITEP-LAT/2002-10}\\
{\large KANAZAWA-02-20}
\end{flushright}
\vspace{1.0cm}}

\begin{document}
\baselineskip=14pt
\begin{center}

\itep

{\large\bf Gluodynamics in external field in dual superconductor
approach}

\vskip 1.0cm
{\bf M.N.~Chernodub}
\vskip 4mm
{\it ITEP, B.~Cheremushkinskaya 25, Moscow, 117259, Russia and} \\
{\it Institute for Theoretical Physics, Kanazawa
University, Kanazawa 920-1192, Japan}

\end{center}

\begin{abstract}
We show that gluodynamics in an external Abelian electromagnetic
field should possess a deconfining phase transition at zero
temperature. Our analytical estimation of the critical external
field is based on the dual superconductor picture which is
formulated in the Euclidean space suitable for lattice
calculations. A dual superconductor model corresponding to the
$SU(2)$ gluodynamics possesses confinement and deconfinement
phases below and, respectively, above the critical field. A dual
superconductor model for the $SU(3)$ gauge theory predicts a rich
phase structure containing confinement, asymmetric confinement and
deconfinement phases. The quark bound states in these phases are
analyzed. Inside the baryon the strings are $Y$--shaped as
predicted by the dual superconductor picture. This shape is
geometrically asymmetric in the asymmetric confinement phases. The
results of the paper can be used to check the dual superconductor
mechanism in gluodynamics.
\end{abstract}

\vskip 0.3cm {\bf 1.} At present there are two popular approaches
to the problem of color confinement in gluodynamics. They are
based on the Abelian monopole~\cite{DualSuperconductor} and on the
center vortex~\cite{CenterVortex} pictures of the gluodynamics
vacuum. In this paper we discuss the Abelian monopole approach
which suggests that confining degrees of freedom of the vacuum in
an Abelian projection~\cite{AbelianProjections} can be described
as a dual superconductor. The key element of this picture is the
monopole condensate which squeezes a chromoelectric flux to a
confining string due to the Meissner effect. The string is an
analog of the Abrikosov vortex~\cite{AbrikosovVortex} in an
ordinary superconductor while the Abelian monopoles are playing
the role of the Cooper pairs. This picture has been confirmed in
many numerical simulations on the lattice (for a review see, {\it
e.g.}, Ref.~\cite{Review}).

Here we investigate the properties of the $SU(2)$ and $SU(3)$
gluodynamics in the external electromagnetic field using the dual
superconductor approach. Our study is motivated by the fact that
the response of the vacuum of a gauge theory on external fields
may provide interesting information about the vacuum structure.
The external fields were used to study nonperturbative properties
of QCD~\cite{QCD-external}, the baryogenesis in the electroweak
theory~\cite{EW-external}, features of various topological
defects in three dimensional models~\cite{ExternalAbelian}, {\it
etc}.

A common feature of known superconductors is the Meissner effect:
the superconductors expel relatively weak external magnetic flux
from their interior. Strong enough magnetic field,
$H^{\mathrm{ext}} \geqslant H_{\mathrm{cr}}$ destroys the superconductivity
and the superconductor goes in the normal (metal) state. In this
paper we estimate analytically the critical electromagnetic fields
which break the dual superconductivity in the $SU(2)$ and $SU(3)$
gluodynamics. Consequently, the confinement is (partially, in the
case of $SU(3)$) lost at these critical fields.

We consider the external electromagnetic fields in a particular
Abelian gauge, which is used to define the Abelian degrees of
freedom. Strictly speaking, these purely Abelian fields are
unlikely to be realized in Nature\footnote{Note, however, that a
general non--Abelian field may have a non--zero projection on the
Abelian subspace discussed in the paper.}. Nevertheless, the
response of the gluodynamics media on these purely Abelian fields
can be used for further checks of the dual superconductivity
hypothesis in numerical simulations of lattice gluodynamics. Since
the dual superconductor picture describes various
non--perturbative phenomena~\cite{Review} such numerical test is
physically motivated.

We consider the gluodynamics at zero temperature because in this
case the relevant couplings of the corresponding dual
superconductor models are already known. Indeed, in the absence of
the external field the gluodynamics experiences the deconfining
phase transition at sufficiently high temperature. On the other
hand, at the critical temperature the dual superconductivity was
demonstrated to be destroyed~\cite{MonopoleCondensation}. Thus the
couplings of the dual superconductor (at least, the value of the
monopole condensate) must depend on the temperature. This
dependence is not known at the time being.

Apart from knowledge of the dual couplings, another simplification
comes from the Euclidean formulation. The action of the four
dimensional dual superconductor is just a trivial dimensional
generalization of the (Helmholtz) free energy functional of the
ordinary three dimensional superconductor. Thus, from the point of
view of {\it static} effects -- such as a response of the
superconductor on a static external field -- the Euclidean dual
superconductor model of the gluodynamics describes just an
infinitely large four dimensional superconducting material. The
electric and magnetic components of the electromagnetic field
differ only by the orientation of the field strength tensor in the
coordinate space with respect to the time axis. However, in the
Euclidean formulation at zero temperature no distinguished
time--direction exists. In this particular respect there is no
difference between external the static electric and the static
magnetic fields (this in no more correct in the presence of the
external sources such as heavy quarks or monopoles). Therefore in
this article we are using the terminology "electromagnetic (EM)
field" for this particular case.

Yet another simplification is due to the fact that the vacuum of
$SU(2)$ and $SU(3)$ Yang--Mills theories in the Abelian projection
is known to be close to the border between type--I and type--II
dual superconductors~\cite{TypeI-II}. At the borderline -- called
also "the Bogomol'ny limit"~\cite{Bogomolny:1975de} -- analytical
results for the string tension are available. Below we consider
the dual models for both theories in the Bogomol'ny limit. We
assume that the external field does not change the couplings of
the dual superconductor model\footnote{This assumption works well
in the macroscopic (Ginzburg--Landau) description of the ordinary
superconductivity.}.

\vskip 0.3cm {\bf 2.} Let us first consider the $SU(2)$ gauge
theory in the $4D$ Euclidean space. The infrared properties of the
vacuum of this model can be described by the Abelian Higgs
(or, Ginzburg--Landau) Lagrangian:
\beqn
\cL_{GL}[B,\Phi] = \frac{1}{4} F^2_{\mu\nu} +
\frac{1}{2} {\bigl| D_\mu(B) \, \Phi\bigr|}^2
+ \lambda {\biggl({|\Phi|}^2 - \eta^2 \biggr)}^2\,,
\label{Lagrangian:SU2}
\eeqn
where $F_{\mu\nu} = \partial_\mu B_\nu - \partial_\nu B_\mu$ is
the field strength for the dual gauge field $B_\mu$, $\Phi$ is the
monopole field with the magnetic charge $g_M$ and $D_\mu =
\partial_\mu + i g_M B_\mu $ is the covariant derivative. The
gauge field $B_\mu$ is dual to the third component of the gluon
field in an Abelian gauge. The model possesses the dual $U(1)$ gauge
symmetry, $B_\mu \to B_\mu - \partial_\mu \alpha$, $\Phi \to e^{i
g_M \alpha}\, \Phi$. The form of the potential implies the
existence of the monopole condensate, $|\langle\Phi\rangle| =
\eta$ with $\eta^2 >0$, and, consequently, non--zero masses of the
dual gauge, $m_B = g \eta$, and monopole, $m_\Phi = 2 \sqrt{2
\lambda} \eta$, fields.

The Bogomol'ny limit corresponds to a region of the coupling space
where the masses of the monopole and gauge fields are the same. In
our notations~\eq{Lagrangian:SU2} the Bogomol'ny limit is defined
by the condition
\beqn
g^2_M \slash \lambda = 8\,.
\label{Bogomolny:SU2}
\eeqn

The properties of the Ginzburg--Landau model~\eq{Lagrangian:SU2}
are very well known. Below we briefly derive the value of the
critical field in the $U(1)$ model~\eq{Lagrangian:SU2} and later
we apply this method to a more complicated ${[U(1)]}^2$ case
corresponding to the $SU(3)$ gluodynamics.

Consider the
four--dimensional sample of the (dual) superconductor occupying
half--space, $x_2 \geqslant 0$. Let us apply the constant external
EM field $F^{\mathrm{ext}}_{\mu\nu} = \varepsilon_{\mu\nu34}
H^{\mathrm{ext}}$ to the boundary of the superconductor. A weak external
field penetrates inside the sample up to the distance $\sim m^{-1}_B$.
The screening of the field (the Meissner effect) is realized due
to the induced superconducting current,
\beqn
J_\mu = \Im m \bigl(\Phi^* D_\mu(B) \Phi\bigr) \equiv |\Phi|^2
\cdot v_\mu\,, \quad  v_\mu = \partial_\mu \varphi + g_M B_\mu\,,
\label{current}
\eeqn
where we have set $\Phi = |\Phi| e^{i \varphi}$. The current is
parallel to the boundary of the superconductor. The monopole
kinetic term in Eq.~\eq{Lagrangian:SU2} can be written as $|D_\mu
\Phi|^2 = (\partial_\mu |\Phi|)^2 + |\Phi|^2 v^2_\mu$. Clearly, a
non--zero current provides an additional positively--defined term
in the Lagrangian ($\propto |\Phi|^2$). As a result, the external
field lowers the value of the monopole condensate. We are
looking for the critical value of the EM field which destroys the
monopole condensate and, consequently, confinement.

We disregard the quantum fluctuations of the fields in the model
\eq{Lagrangian:SU2} treating this system classically. To derive
the critical EM field it is convenient to rewrite the action of
the model~\eq{Lagrangian:SU2} as a two--dimensional integral. The
first direction of the two-dimensional plane is obviously the
depth of the dual superconductor, $x_2$,
while the second is given by the direction of the
current~\eq{current}. We choose $J_\mu \propto \delta_{\mu,1}$ and
rewrite the model~\eq{Lagrangian:SU2} in the $(1,2)$ plane. The
first term of eq.~\eq{Lagrangian:SU2} is $F^2_{\mu\nu}/4  = H^2/2$
while the second term can be rewritten with the help of relations:
\beqn
{\bigl|D_\mu \Phi\bigr|}^2 = \sum\limits_{\alpha=1,2}
{\bigl|D_\alpha \Phi\bigr|}^2 =
{\biggl| \biggl(D_1 \pm i D_2 \biggr)\Phi\biggr|}^2
\mp 2 \varepsilon_{\alpha\beta} \partial_\alpha J_\beta \pm
i \Phi^* [D_1,D_2 ] \Phi\,,
\label{twostars}
\eeqn
The last term in this equation can also be represented as
$i \Phi^* [D_1,D_2 ]  \Phi = - g_M H \, {|\Phi|}^2$. We get the
following action in the $(1,2)$ plane:
\beqn
S_{GL} = L_3 L_4 \int \dd^2 x \Bigl[\frac{1}{2} H^2 +
\frac{1}{2} {\Bigl| \Bigl(D_1 \pm i D_2 \Bigr)\Phi\Bigr|}^2
- \frac{g_M}{2} H \, {|\Phi|}^2
+ \lambda {\Bigl({|\Phi|}^2 - \eta^2 \Bigr)}^2\Bigr] + S_J\,,
\label{Lagrangian:SU2:2}
\eeqn
where $L_i$ is the (infinite large) length of the dual
superconductor in $i$th direction and $S_J$ is the action of the
surface current:
\beqn
S_J = L_3 L_4 \int \dd^2 x\, \varepsilon_{\alpha\beta} \partial_\alpha
J_\beta = L_1 L_3 L_4 \cdot J_1(x_2=0)\,.
\eeqn
Using condition \eq{Bogomolny:SU2} we rewrite
Eq.~\eq{Lagrangian:SU2:2} further:
\beqn
S = \frac{1}{2} L_3 L_4 \int \dd^2 x \Bigl[
{\biggl| \biggl(D_1 \pm i D_2 \biggr)\Phi\biggr|}^2 +
{\biggl[H \mp \frac{g_M}{2} \biggl(|\Phi|^2 - \eta^2
\biggr)\Bigr]}^2\biggr] \mp S_{flux} + S_J\,.
\label{Lagrangian:SU2:3}
\eeqn
The sign in this equation is dictated by the "flux"
action, $S_{flux} = g_M L_3 L_4 \eta^2 \int \dd^2 x\, H \slash 2$,
which must be positive. We choose $H>0$ and get two Bogomol'ny equations
which minimize the action~\eq{Lagrangian:SU2:3}:
\beqn
(D_1 - D_2) \, \Phi = 0\,,\quad H + \frac{g_M}{2}\,
\Bigl(|\Phi|^2 - \eta^2\Bigr) = 0\,.
\nonumber
\eeqn
The second equation gives the value of the
monopole condensate at the boundary
\beqn
|\Phi(x_2=0)|^2 = \eta^2 - 2 H^{ext} \slash g_M\,.
\label{Phi:x2=0}
\eeqn
The condensate disappears at the critical value of the EM field,
$H^{\mathrm{ext}}= H_{\mathrm{cr}} = g_M \eta^2 \slash 2$. At this value the
superconducting current at the boundary vanishes, $J_1=0$, giving
$S_J=0$. The total action is given by the flux contribution, $S =
S_{flux} = Vol \cdot g^2_M \eta^4 \slash 4$ which is the sum of the
free energy of the normal state, $S_n\equiv S_{GL}[B=0,\Phi=0]$
and the free energy of the EM field, $S_H$: $S_n=S_H = Vol \cdot
g^2_M \eta^4 \slash 8$.

In the Bogomol'ny limit the tension of the string spanned on
trajectories of the fundamental charges can be evaluated
exactly\footnote{We stress that this result is obtained at zero
temperature in the absence of the external
fields.}~\cite{Bogomolny:1975de,VeSc76}, $\sigma=\pi \eta^2$.
Using the Dirac relation between magnetic ($g_M$)
and electric ($g$) charges, $g_M g = 2 \pi$, we get the exact
value of the critical EM field in terms of the string tension:
\beqn
g H_{\mathrm{cr}} \slash \sigma = 1\,, \qquad \mbox{for SU(2)}\,.
\label{Hcr:SU2}
\eeqn
This equation can be used in numerical simulations to check
independently the closeness of the gluodynamics vacuum to the
boundary between type--I and type--II superconductors.

The nature of the phase transition at the critical EM field can be
understood as follows. The EM field, applied to the boundary of
the media with the $SU(2)$ gauge fields, lowers the value of the
monopole condensate according to Eq.~\eq{Phi:x2=0}. For the
external fields $H^{\mathrm{ext}} < H_{\mathrm{cr}}$ the monopole
condensate is non-zero, $|\Phi(x_2=0)| > 0$. This implies, that
the dual photon mass at the boundary is non-zero as well,
$m_B(x_2=0) = g_M |\Phi(x_2=0)|$. Due to the Meissner effect
the external field diminishes as we go deeper into the media. At
the distances of the order of the correlation length, $\sim
m^{-1}_\Phi$, the monopole condensate restores at bulk value,
$\Phi \sim \eta$. Consequently, at these distances the dual photon
mass is restored as well, $m_B \sim g_M \eta$ and the external EM
field diminishes exponentially, $H(x_2) \sim e^{- m_B\, x_2}$ for
$x_2 \gtrsim m^{-1}_\Phi$. Thus, no EM field is present in the
bulk of the system.

Now suppose that the external field reaches its critical value,
$H^{\mathrm{ext}} = H_{\mathrm{cr}}$. Then both the monopole
condensate and the dual photon mass are zero at the boundary. The
last fact implies that the field penetrates inside the media for
an infinitesimally small distance $\delta x_2$ without any
suppression because the Meissner effect is absent. At the distance
$\delta x_2$ the situation repeats again: the condensate and the
photon mass are zero and the field penetrates deeper into the media for
another infinitesimal step, {\it etc}. Thus, the external field
"eats" the condensate step by step and finally the condensate
disappears in the whole space. Due to this mechanism the EM field
originated at the boundary can destroy the monopole condensate
(and, consequently, the confinement) in the bulk.

\vskip 0.3cm {\bf 3.}
Now let us consider the Lagrangian of ${[U(1)]}^2$ Higgs model
corresponding to $SU(3)$ gluodynamics~\cite{MaSu81}:
\beqn
\cL = \frac{1}{4} F^a_{\mu\nu} F^{a,\mu\nu} +
\sum\limits^3_{i=1} \Bigl[\frac{1}{2}
{\bigl| D^{(i)}_\mu \Phi_i\bigr|}^2 + \lambda
{\Bigl({|\Phi_i|}^2 - \eta^2 \Bigr)}^2\Bigr]\,,
\label{Lagrangian:SU3}
\eeqn
where $F^a_{\mu\nu} = \partial_\mu B^a_\nu - \partial_\nu B^a_\mu$
is the field strength for the gauge fields $B^a_\mu$, $a=3,8$,
$D^{(i)}_\mu = \partial_\mu + i g_M \varepsilon^a_i B^a_\mu$ is the
covariant derivative acting on the monopole fields $\Phi_i$,
$i=1,2,3$. The $\epsilon$'s are the root vectors of the group
$SU(3)$:  $\vec \epsilon_1=(1, 0)\,,\vec \epsilon_2=(-1 \slash 2,
-{\sqrt{3} \slash 2})\,,\vec \epsilon_3=(-{1 \slash 2}, {\sqrt{3}
\slash 2})$. No summation over the Latin index $i$ is implied.

The gauge fields $B^{3,8}_\mu$ are dual to the diagonal components
$a=3,8$ of the gluon field $A^a_\mu$. Lagrangian
\eq{Lagrangian:SU3} respects the dual $[U(1)]^2$ gauge invariance:
$B^a_\mu \to B^a_\mu + \partial_\mu \alpha^a$, $\theta_i \to
\theta_i + g_M (\varepsilon^3_i \alpha^3 + \varepsilon^8_i
\alpha^8)$, $a=3,8$, $i =1,2,3$, where $\alpha^3$ and $\alpha^8$
are the parameters of the gauge transformation. The phases of the
monopole fields satisfy the relation
\beqn
\sum\nolimits^3_{i=1} \arg \Phi_i =0\,,
\label{phases:Phi}
\eeqn
which plays an important role in the formation of the quark
bound states within the dual superconductor formalism~\cite{BoundStates,KoCh}.

The Bogomol'ny limit is defined by condition~\cite{Chernodub:1999xi,Weyl1}
\beqn
g^2_M \slash \lambda = 16 \slash 3\,,
\label{Bogomolny:SU3}
\eeqn
and the equations of motion in this limit are given by
\beqn
\Bigl(D^{(i)}_1 \pm i D^{(i)}_2 \Bigr)  \chi_i = 0\,, \quad
H^{(i)} \mp \frac{3 g_M}{4} \Bigl({|\chi_i|}^2 - \eta^2
\Bigr) = 0\,;\quad i =1,2,3\,,
\label{EOM}
\eeqn
where
\beqn
H^{(i)} = \sum_{a=3,8} \epsilon^a_i H^a\,,
\label{Hi:def}
\eeqn
are the EM fields projected on the $(3,8)$-charges of the monopole
fields. One can imagine these three fields as the {\it dual} red,
{\it dual} blue and {\it dual} green EM fields.

The second equation in \eq{EOM} gives the same critical value
for all components of the $H^{(i)}$--fields:
\beqn
H^{(i)}_{\mathrm{cr}} \equiv {\tilde H}_{\mathrm{cr}} = 3 g_M \eta^2 \slash 4\,.
\label{criticality:SU3}
\eeqn
The critical values are equivalent due to the Weyl symmetry of the
dual model~\cite{Weyl1,Weyl2} which states that the ${[U(1)]}^2$
Lagrangian~\eq{Lagrangian:SU3} is invariant under the
transformations of the dual gauge fields $B^{3,8}$
corresponding to the mutual permutations of the $H^{(i)}$
fields~\eq{Hi:def}.

\begin{figure*}[!htb]
  \begin{center}
      \epsfxsize=12.0cm \epsffile{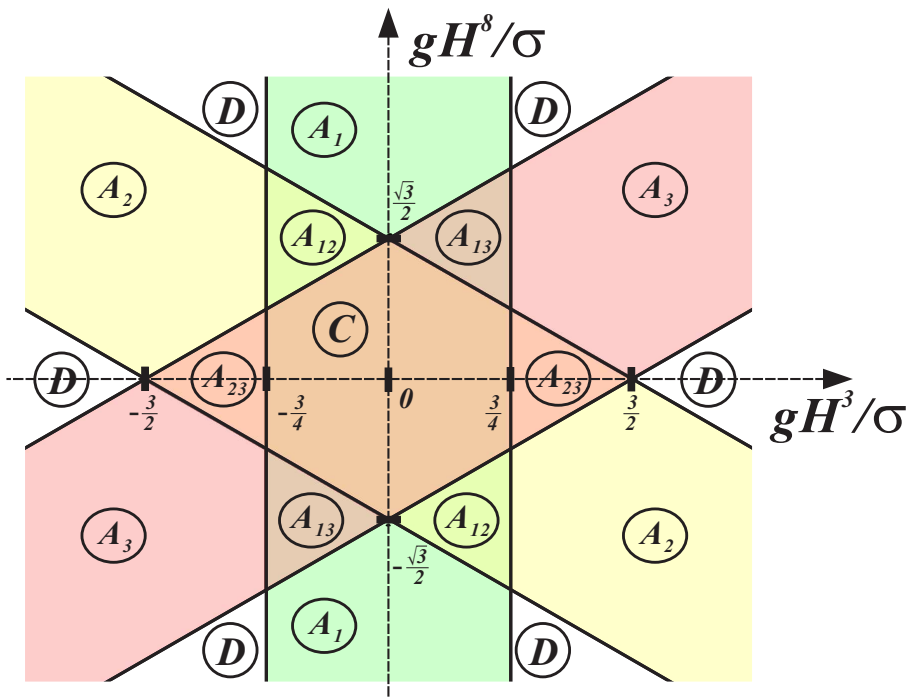}
  \end{center}
  \vspace{-0.5cm}
  \caption{The zero temperature
  phase diagram of the dual ${[U(1)]}^2$ Higgs model in the Bogomol'ny limit
  in the presence of the external electromagnetic field.
  The confinement and deconfinement
  phases are denoted as $C$ and $D$. The asymmetric confinement
  phases are denoted as $A_{i}$ and $A_{jk}$, where subscripts
  indicate the components of the monopole field $\Phi_i$ which are
  condensed in these phases.}
  \label{fig:phase}
\end{figure*}

The string tension spanned between the fundamental charges
(quarks) in the Bogomol'ny limit~\eq{Bogomolny:SU3} of the $SU(3)$
gluodynamics is~\cite{Chernodub:1999xi,Weyl1} $\sigma = 2 \pi
\eta^2$. Using the Dirac quantization condition we get the
critical field ${\tilde{H}}_{cr}$ in units of the string tension:
\beqn
g {\tilde{H}}_{cr} \slash \sigma = 3 \slash 4\,,
\qquad \mbox{for SU(3)}\,.
\label{Hcr:SU3}
\eeqn
When the strength of the EM component $H^{(i)}$ reaches the
${\tilde H}_{cr}$ value then the expectation value of the
corresponding component of the monopole field, $\Phi_i$, gets
vanished. Note, however, that the fields $H^{(i)}$ play an
auxiliary role because they are not independent according to
Eq.~\eq{Hi:def}. Expressing the auxiliary fields $H^{(i)}$ in
terms of the components the EM field, $H^{3,8}$, and using
Eqs.(\ref{criticality:SU3},\ref{Hcr:SU3}) we get the phase diagram
depicted in Figure~\ref{fig:phase}.

The phase diagram contains confinement ($C$), deconfinement ($D$)
and the asymmetric confinement phases ($A$). The position of the phase
transition depends not only on the absolute value of the EM field
but also on the ("color") orientation of this field in the Cartan
subgroup. At low values of the field the model is always confining
regardless of the color orientation. However, as the absolute
value of the field is increased, the model enters -- depending on
color orientation -- one of six ($A_{12}$, $A_{13}$, $A_{23}$,
$A_{1}$, $A_{2}$ or $A_{3}$) asymmetric confinement phases. In the
$A_{ij}$ phase the $i$th and $j$th components of the monopole
field are condensed while the expectation value of the third
component is zero. In the phase $A_i$ the $i$th component is
condensed while the others two components are not. With the
further increase of the field the model either enters the
deconfinement phase, $D$, or stays in one of the three
asymmetric confinement phases, $A_{1}$, $A_{2}$ or $A_{3}$.

\vskip 0.3cm {\bf 4.} The quark bound states in the Abelian
projection approach are classified with the help of the states of
the strings spanned between the constituent quarks. The string
configuration in the baryon was extensively studied both in the
Abelian projection approach~\cite{BoundStates,KoCh} as well as in
a gauge independent formalism~\cite{MitrubishiShape}. The dual
superconductor model predicts~\cite{KoCh} the existence of the
$Y$--shaped string configuration in agreement with most of
Ref.~\cite{MitrubishiShape}. Here we discuss the quark bound
states in the presence of the external Abelian field.

The confining properties of both the confinement and the
deconfinement phases are standard. In the confinement phase all
three monopole fields, $\Phi_i$, $i=1,2,3$, are condensed and the
$[U(1)]^2$ model possesses three types of vortex
solutions~\cite{KoCh,Weyl2,An} which confine all quarks into bound
states. Each vortex solution is characterized by the winding
number $n_i$, $\Phi_i \propto e^{i n_i \phi_i}$, where $\phi_i$ is
the azimuth angle. The winding numbers of the strings are
subjected to the constraint $n_1 + n_2 + n_3 =0$ coming from
Eq.~\eq{phases:Phi}. In the deconfinement phase all monopole
fields are not condensed and the bound stated are not
formed\footnote{In this paper we disregard weakly bounded states
which might appear due to the exchange of the perturbative gluons
between quarks. We also disregard the role of the
Jacobian~\cite{Jacobian} arising in the string model.}.

Let us discuss the bound states in the asymmetric confinement phases.
Consider, for example, $A_{12}$ phase. In this case the model
possesses only two condensates, $\langle \Phi_{1,2}\rangle \neq
0$. Consequently, on the classical level only two types of the
chromoelectric strings can be formed with $n_{1,2} \neq 0$.
Despite the third field has zero expectation value, $\langle
\Phi_i \rangle =0$, its phase may fluctuate and the (classically
tensionless) $n_3 \neq 0$ string--like configurations may appear.
This implies that in addition to the {\it purely classical} $\vec
n = (1,-1,0)$ string configuration there exist also $(1,0,-1)$ and
$(0,1,-1)$ configurations. These additional configurations are
composed from the classical strings with either $n_1 \neq 0$ or
$n_2 \neq 0$ and a string--like (tensionless) quantum excitation
with $n_3\neq 0$. The stability of the quantum components of such
configurations is guaranteed by Eq.~\eq{phases:Phi}.

The $\vec n = (1,-1,0)$ string configuration must have bigger
string tension than $(-1,0,1)$ and $(0,1,-1)$ configurations. Thus
in the $A_{12}$ phase we expect the existence of the relatively
heavy meson composed from red quarks and two lighter mesons made
of blue and green quarks\footnote{Here we adopt the classification
of Ref.~\cite{KoCh} assuming that the quarks in a meson state are
connected with each other by a pair of the strings with winding
numbers $\vec n = (1,-1,0)$, $(-1,0,1)$ and $(0,1,-1)$ for $R\bar
R$, $B \bar B$ and $G\bar G$ states respectively.}. The observed
color asymmetry in the meson states is caused by the breaking of
the Weyl symmetry~\cite{Weyl1,Weyl2} of the
Lagrangian~\eq{Lagrangian:SU3} by the external field.

\begin{figure*}[!htb]
  \begin{center}
      \epsfxsize=13.0cm \epsffile{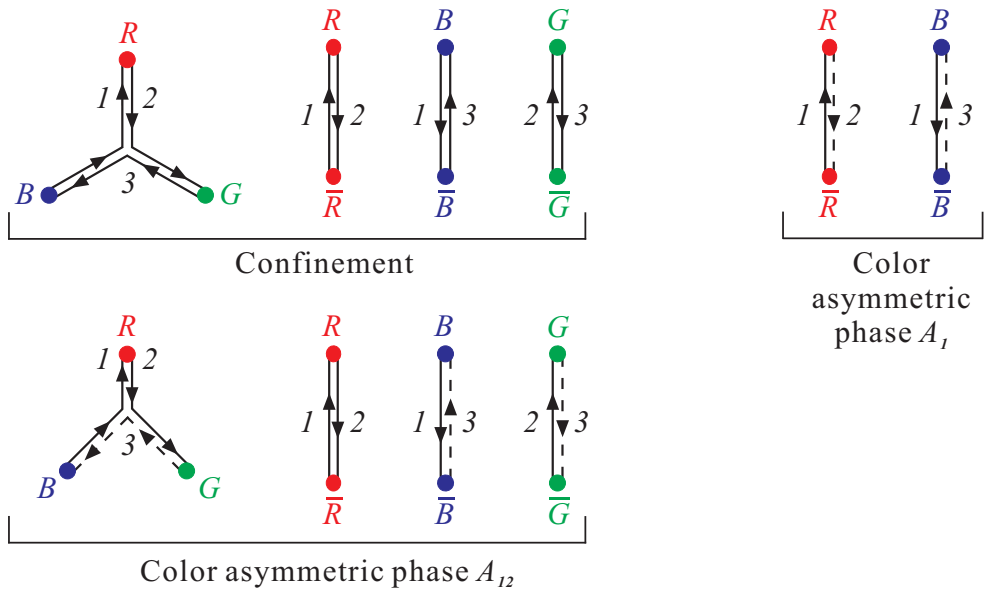}
  \end{center}
  \vspace{-0.5cm}
\caption{The meson and baryon bound states in confinement,
deconfinement and $A_{12}$, $A_1$ asymmetric confinement phases.
The solid lines correspond to the strings existing on the classical
level while the dashed lines represent the quantum string excitations.}
  \label{fig:states}
\end{figure*}

The baryon state also exists in the $A_{12}$ phase but it should
be lighter than the baryon in the confinement phase. The quarks in
the baryon are connected to each other by all three types of the
string configurations. It is known~\cite{KoCh} that in the absence
of the external field these strings form a symmetric $Y$--shaped
profile. In the $A_{12}$ phase the tension of the $(1,-1,0)$
string configuration is heavier than that of $(-1,0,1)$ and
$(0,1,-1)$ configurations. Therefore in this phase the strings
in the baryon state must form an {\it asymmetric} $Y$--profile, see
Figure~\ref{fig:states}.

The properties of the $A_{i}$ phases are different from those of
the $A_{ij}$ phases. In particular, in the $A_1$ phase the string
with the winding number $n_1 \neq 0$ has a non--zero string
tension while the strings with $n_{2,3}\neq 0$ are classically
tensionless. Thus in this phase only $(1,-1,0)$ and $(1,0,-1)$
string configurations have a non--zero string tension. This
implies the existence of the mesons composed of red and blue
quarks while the meson made of green quarks is absent. The
existing mesons must be lighter than those in the confinement
phase. All possible meson and baryon states in confinement,
deconfinement and asymmetric confinement phases are depicted in
Figure~\ref{fig:states}.

\vskip 0.3cm {\bf 5.} Summarizing, we have explored the phase
structure of the $SU(2)$ and $SU(3)$ gauge theories in the
external electromagnetic fields at zero temperature. Both theories
are considered in the dual superconductor formalism formulated in
the Euclidean space (suitable for lattice calculations) and in the
Bogomol'ny limit (as confirmed by various lattice calculations).

We have found that the phase diagram of the dual superconductor
model corresponding to the $SU(2)$ gluodynamics contains
confinement and deconfinement phases which are located below and,
respectively, above the critical field. The critical
electromagnetic field is analytically estimated in terms of the
string tension in Eq.~\eq{Hcr:SU2}.

The phase diagram of the ${[U(1)]}^2$ dual superconductor
corresponding to the $SU(3)$ gluodynamics contains 8 phases
(confinement, deconfinement and 6 asymmetric confinement phases).
The reach phase structure -- shown in Figure~\ref{fig:phase} --
comes from the dependence of the monopole condensate on the color
orientation of the external field. This finding is supported by
the observation of Ref.~\cite{DifferentMonopolesSU3} that the
properties of the Abelian monopoles depend on the color
orientation of the monopole. Three of six asymmetric confinement
phases ($A_{12}$, $A_{13}$ and $A_{23}$) contain one baryon and
three meson states. Two of the meson states are lighter then the
third one. The strings in the baryon form the asymmetric
$Y$--shaped profile. These phases can be still regarded as
confinement phases since the quarks of all three colors are
confined.

The other three asymmetric confinement phases ($A_1$, $A_2$ and $A_3$) may
contain only two light meson states while the baryon state is
absent at all. The quarks carrying a particular (phase--dependent)
color are not confined in these phases.

Note, that our results were obtained for the Abelian external
fields which are applied in a {\it fixed} Abelian
projection\footnote{Here we do not discuss a controversy topic of
the (in)dependence of the dual superconductor picture on chosen
Abelian projection~\cite{Controversy}. We assume that this picture
is realized in the Maximal Abelian projection, see
Ref.~\cite{Review} for a review.}. In particular, this means that
the recent results of Refs.~\cite{CeaCosmaiRecent} for the phase
diagram in the external Abelian fields can not be compared with
our predictions because these results were obtained without the
gauge fixing.

It would be interesting to check the predictions of this paper by
numerical simulations performed in the Maximal Abelian projection
of the $SU(2)$ and $SU(3)$ gluodynamics. As it is shown above the
dual superconductor hypothesis predicts a particular phase diagram
in the external electromagnetic fields. The numerical
investigation of this diagram can be used for further checks of
the dual superconductivity of the vacuum. Moreover, the value of
the critical fields corresponding to the phase transitions (if
exist) could be used to determine the closeness of parameters of
the dual superconductor to the type--I/II boundary.

\section*{Acknowledgments}
The author is grateful to V.G.~Bornyakov and Y.~Koma for
interesting discussions. The work is supported by the JSPS
Fellowship P01023.

\end{document}